# Observation of Coherent Perfect Acoustic Absorption at an Exceptional Point

Yi-Fei Xia[1,*], Zi-Xiang Xu[1,*], Yu-Ting Yan[1,*], An Chen[1], Jing Yang[1,†], Bin Liang[1,‡], Jian-Chun Cheng[1,∥], and Johan Christensen[2,§]

[1]Collaborative Innovation Center of Advanced Microstructures and Key Laboratory of Modern Acoustics, MOE, Institute of Acoustics, Department of Physics, Nanjing University, Nanjing 210093, People's Republic of China.

[2]IMDEA Materials Institute, Calle Eric Kandel, 2, 289006, Getafe, Madrid, Spanish.

[*]These three authors contributed equally to this work.

Corresponding to: [†]yangj@nju.edu.cn, [‡]liangbin@nju.edu.cn, [∥]jccheng@nju.edu.cn, and [§]johan.christensen@imdea.org.

Non-Hermitian systems have recently shown new possibilities to manipulate wave scattering by exploiting loss, yet coherent perfect absorption at an exceptional point (CPA EP) remains elusive in acoustics. Here we demonstrate it based on a two-channel waveguide with compact lossy resonators. We realize imbalanced losses crucial for CPA EP by using active components to independently modulate the non-Hermiticity. The CPA EP experimentally manifests as full absorption at a unique real frequency and shows high sensitivity to the incident phase variations. Our findings open an avenue to explore novel non-Hermitian physics for classical waves and develop innovative acoustic singularity-based devices.

***Introduction.*** - The Hermiticity of a Hamiltonian ensures the conservation of energy and shapes the physical reality in many systems [1,2]. However, in non-conservative systems, interaction with the environment leads to non-Hermitian dynamics. The past decade has witnessed a surge of research on non-Hermitian physics [3-5], resulting in unprecedented principles, phenomena, and applications in both quantum and classical systems [6,7]. By tailoring gain and loss, non-Hermitian systems exhibit intriguing phenomena near exceptional points (EPs), which are singularities in the parameter space where multiple eigenvalues and the associated eigenstates coalesce [8-15]. This coalescence is accompanied by a plethora of exotic phenomena, such as the skin effect [16-19], chiral state transfer [20,21], and non-Abelian braiding [22]. Moreover, the introduction of non-Hermiticity has given rise to various applications related to EPs in open wave systems, including unidirectional invisibility, single-mode lasing, and enhanced sensing [23-25].

As a typical non-Hermitian effect in wave physics, coherent perfect absorption (CPA) occurs under purely incoming boundary conditions when the zeros of the scattering matrix lie on the real frequency axis [26], which has attracted rapidly growing attention in the past few years and been extended into diverse fields including acoustics and mechanical waves [27], microwaves [28], and optical waves [29,30]. Thanks to the intrinsic stability of CPA ensured by the wave interference that entraps the incident radiation inside the lossy media, it provides the possibility that two purely incoming solutions coalesce at a unique real frequency [31], referred to as CPA EPs, which is recently predicted and observed in optics [32]. This offers a new way of broadening absorbing bandwidth [33] and enhancing sensitivity [22]. Limited by the macroscopic wavelength of sound waves and difficulty in modulating acoustic non-Hermiticity to achieve imbalanced losses, however, realizing CPA EP in non-Hermitian acoustic scattering systems with compactness and simplicity so far remains elusive.

Here, we propose an acoustic non-Hermitian scattering system consisting of two-channel waveguides coupled to imbalanced lossy resonant cavities to achieve CPA EP. Based on the coupled-mode theory, we analytically derive the scattering matrix of the system and predict the critical conditions for CPA EP. Furthermore, we introduce a metamaterial-based implementation that is significantly smaller in size compared to its optical counterparts, with coupling strengths and intrinsic losses independently tunable through active acoustic units. We experimentally observe the occurrence of CPA EP in this system, showing the expected strong absorption at the single resonant frequency and distinctive sensitivity to the phase variations of incoming acoustic waves, which is consistent with theoretical and simulation results. Our work provides deeper insight to the non-Hermitian physics in classical wave systems and may open avenues to the design and practical application of novel acoustic absorbers, sensors, and directional devices, etc.

***Acoustic non-Hermitian scattering system for realizing CPA EP.*** - To realize CPA EP, we propose an acoustic non-Hermitian scattering system, as illustrated in Figure 1(a), where two coherent acoustic plane waves normally impinge on both sides of the system along opposite propagation channels. These incoming acoustic waves induce lossy resonance in the double cavities, resulting in destructive interference effects that trap and dissipate the radiation within the acoustic scattering system. By adjusting the non-Hermiticity of the system, the resonant modes in the cavities could become degenerate, leading to the emergence of CPA EP, which manifests as perfect absorption at a real frequency. Specifically, the absorption behavior of such a system is described by a scattering matrix $S$ with reflection and transmission coefficients. Then the amplitudes of incoming and outgoing acoustic waves can be related by $S$ as

$$\begin{pmatrix} p_{o1} \\ p_{o2} \end{pmatrix} = S \begin{pmatrix} p_{i1} \\ p_{i2} \end{pmatrix} = \begin{pmatrix} r_1 & t_2 \\ t_1 & r_2 \end{pmatrix} \begin{pmatrix} p_{i1} \\ p_{i2} \end{pmatrix}, \quad (1)$$

where $p_{i1,2}$ and $p_{o1,2}$ respectively refer to the acoustic pressure for the incoming and outgoing ones, $r_{1,2}$ and $t_{1,2}$ signify the acoustic reflection and transmission coefficients respectively, and the subscripts 1 and 2 represent the different acoustic propagating channels for such an acoustic non-Hermitian scattering system.

The scattering characteristics of this system depend on the resonance frequency, coupling strength, and intrinsic losses of the two cavities, which can be described using a second-order non-Hermitian Hamiltonian [34]:

$$H = \begin{pmatrix} \omega_1 - i(\gamma_1 + \gamma_{c1}) & \kappa \\ \kappa & \omega_2 - i(\gamma_2 + \gamma_{c2}) \end{pmatrix}. \quad (2)$$

Here, $\omega_{1,2}$ are the inherent resonant frequencies of the cavities in the absence of any coupling or losses. For compact subwavelength cavities, their resonance frequencies are primarily determined by their dimensions and geometries. $\kappa$ denotes the internal coupling strength, reflecting the transfer of acoustic energy between the two cavities. $\gamma_1$ and $\gamma_2$ represent the intrinsic losses of the cavities resulting from acoustic dissipation mechanisms, which can be adjusted using active acoustic units. $\gamma_{c1}$ and $\gamma_{c2}$ denote the external losses between the cavities and their corresponding external channels, describing how acoustic energy leaks out from each cavity into the channels.

Based on the coupled-mode theory [35], the scattering matrix $S$ can be expressed as follows:

$$S = 1 - iK^\dagger \frac{1}{\omega - H} K, \tag{3}$$

where $K = \text{diag}(\sqrt{2\gamma_{c1}}, \sqrt{2\gamma_{c2}})$ is the coupling operator, and $\omega$ is the frequency of incident monochromatic acoustic wave [36]. This expression explicitly describes the relationship between the scattering matrix and the system Hamiltonian. Specifically, when $\gamma_{c1} = \gamma_{c2}$, the scattering matrix and the Hamiltonian share the same eigenvectors, and the corresponding eigenvalues satisfy the relationship as

$$\lambda_S = 1 - i \frac{2\gamma_{c1}}{\omega - \lambda_H}, \tag{4}$$

where $\lambda_S$ and $\lambda_H$ are the eigenvalues of the scattering matrix and the Hamiltonian, respectively. When the eigenvalue of the scattering matrix is zero, i.e. $\lambda_S = 0$, the system eliminates the outgoing acoustic waves and achieves CPA. According to Equation (4), the frequency at which CPA occurs is closely related to the eigenvalue of the Hamiltonian, given by $\omega_z = \lambda_H + 2i\gamma_{c1}$. For simplicity while without losing generality, we consider a special case where the two resonant frequencies are equal, i.e., $\omega_1 = \omega_2 = \omega_0$. In this case, the frequency at which CPA occurs can be calculated as follows:

$$\omega_{z1,2} = \omega_0 + i\frac{\gamma_{c1} + \gamma_{c2} - \gamma_1 - \gamma_2}{2} \pm \frac{1}{2}\sqrt{\left(i(\gamma_{c1} - \gamma_1 - \gamma_{c2} + \gamma_2)\right)^2 + 4\kappa^2}. \tag{5}$$

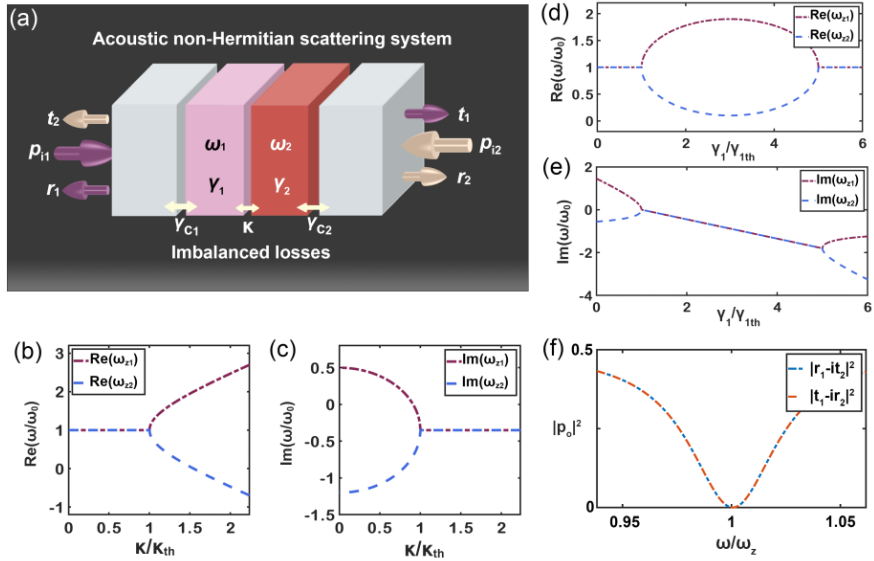

FIG.1. (a) A schematic diagram of the acoustic non-Hermitian scattering system composed of two-side acoustic propagating channels connected by two coupled resonant modes. The two grey regions represent two propagating channels, and the pink and red parts describe the acoustic

resonant modes with imbalanced losses separately. The two-way arrows depict the coupling between acoustic resonant states or between acoustic resonant state and propagation channel. (b) and (c) Phase transition diagrams for the real and imaginary parts of $\omega_{z1,2}$ as functions of normalized coupling strength $\kappa/\kappa_{th}$, respectively. (d) and (e) Phase transition diagrams for the real and imaginary parts of the two eigenvalues passing through a CPA EP as functions of the normalized intrinsic loss $\gamma_1/\gamma_{1th}$. (f) Outgoing acoustic intensity in the two channels varied with frequency. The outgoing acoustic intensity reaches its minimum value of zero when $\omega = \omega_z$, indicating the occurrence of a CPA EP.

In general, there exist two distinct frequencies $\omega_{z1,2}$ both satisfying the CPA conditions, corresponding to the two zeros of the transfer function [36]. However, what is noteworthy is that when the eigenvalues of the Hamiltonian and the scattering matrix simultaneously degenerate, these two zeros merge at the real frequency, resulting in the occurrence of CPA EP. Equation (5) gives the critical condition for CPA EP, expressed as $\gamma_1 + \gamma_2 = \gamma_{c1} + \gamma_{c2}$ and $\kappa = |\gamma_1 - \gamma_{c1}|$ [36]. The phase transition diagrams for the real and imaginary parts of the two solutions $\omega_{z1,2}$ are separately demonstrated in Figure 1(b) and Figure 1(c), as functions of the normalized coupling strength $\kappa/\kappa_{th}$. With the increase of $\kappa$, the real parts of $\omega_{z1,2}$ keep degenerate until $\kappa = \kappa_{th}$, and in the regime of $\kappa > \kappa_{th}$, they become divided and are located on either side of $\text{Re}(\omega) = \omega_0$. Opposite to the dependence of the real parts on the coupling strength $\kappa$, the imaginary parts of $\omega_{z1,2}$ separate at the beginning and coalesce when $\kappa = \kappa_{th}$. In the following, we investigate the degenerate behavior of $\omega_{z1,2}$ as $\gamma_1$ varies and plot typical results in Figure 1(d) and Figure 1(e). When gradually increasing the intrinsic loss $\gamma_1$, we can find two EPs at $\omega_1 \approx \omega_0$ and $\omega_2 \approx \omega_0(1 - i1.8)$. The former is a CPA EP where the zeros are at a real frequency, while the latter is a general EP which cannot be observed experimentally due to being located at a complex frequency. Similarly, the above trend of zero merging can also be observed in phase transition diagrams when tuning other system parameters, such as intrinsic losses $\gamma_{1,2}$ or coupling strengths $\gamma_{c1,2}$.

In the case of CPA EP, both eigenvalues of the scattering matrix are zero, corresponding to the same eigenvector $[1, -i]^T$ [36]. This implies that when the incoming acoustic waves matches this eigenvector, the intensity of the outgoing sound waves reaches its minimum value of zero at frequency $\omega = \omega_z$, as shown in Figure 1(f). At this point, the scattering matrix must take the following form:

$$S = C \begin{pmatrix} i & 1 \\ 1 & -i \end{pmatrix}, \tag{6}$$

where $C$ is a constant. Notably, the scattering matrix is a nilpotent matrix ($S^2 = 0$), which implies that complete absorption can be achieved by cascading two such systems. For general CPA, the phases of the outgoing acoustic waves vary with that of the incoming acoustic waves. For CPA EP, however, the phases of the incoming waves only affect the intensity of the outgoing acoustic waves. When the incoming acoustic waves in the two channels have equal amplitudes but a phase difference of $\Delta\varphi$, the intensities of the outgoing acoustic waves depend on $\Delta\varphi$ through the relationship $|p_o| = |\exp(j\Delta\varphi + \pi/2) + 1|$, and are therefore highly sensitive to phase variations.

***Metamaterial-based acoustic implementation.*** - To implement the acoustic non-Hermitian scattering system, we employ a metamaterial-based approach to construct an acoustic cavity-tube model with subwavelength dimension. As illustrated in Figure 2(a), the proposed model is composed of two-channel waveguides linked to coupled acoustic resonant cavities. This model enables independent modulation of each system parameter in the effective Hamiltonian given in Equation (2), allowing CPA EP to occur at unique real frequency. Specifically, the resonant frequency $\omega_0$ of the two identical cavities coupled with each other can be modulated by designing the inner geometries of cavities. These two compact resonant cavities are connected by a small tube to implement the inter-cavity coupling strength $\kappa$. Two single-mode waveguides for sound propagation are linked to the cavities by two small tubes with the coupling strengths being $\gamma_{c1}$ and $\gamma_{c2}$, respectively. Notice that the coupling strength is inversely proportional to the length of the connecting tube. The intrinsic losses $\gamma_{1,2}$ of the two cavities are modified by controlling the attenuation of sound waves inside the cavities which is realized in simulation by adding a modifiable imaginary part to the speed of sound and experimentally implemented via producing the sound interference effect.

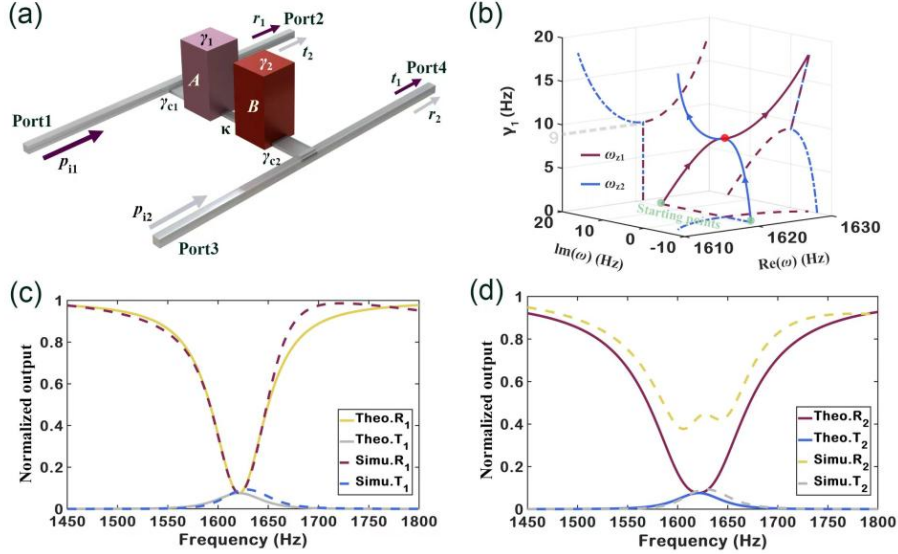

FIG.2. (a) Schematic diagram of the practical implementation which is realized with coupled acoustic resonant cavities connected with double rectangular waveguides as the input and output channels. (b) Trajectories of the eigenvalues of $S$ as $\gamma_1$ varies. (c) Theoretical and simulated reflection spectra $R_1 = |r_1|^2$ and transmission spectra $T_1 = |t_1|^2$ by exciting port 1 only when $\gamma_1 = 9\text{ Hz}$, $\gamma_2 = 31\text{ Hz}$, $\gamma_{c1} = \gamma_{c2} = 20\text{ Hz}$, $\kappa = 11\text{ Hz}$, and $\omega_0 = 1621\text{ Hz}$. (d) Theoretical and simulated reflection spectra $R_2 = |r_2|^2$ and transmission spectra $T_2 = |t_2|^2$ by exciting port 3 only.

Here, we select the system parameters as $\gamma_2 = 31\text{ Hz}$, $\gamma_{c1} = \gamma_{c2} = 20\text{ Hz}$, $\kappa = 11\text{ Hz}$, and $\omega_0 = 1621\text{ Hz}$, and analyze the influence of the variable parameter $\gamma_1$ on the degeneracy of the zeros. The trajectories of eigenvalues $\omega_{z1,2}$ under the variation of $\gamma_1$ are illustrated in Figure 2(b). Initially, the two complex eigenvalues are separately distributed at two starting points and then approach each other as $\gamma_1$ increases. Under the above derived conditions for CPA EP ($\gamma_{c1} + \gamma_{c2} = \gamma_1 + \gamma_2$ and $\kappa = |\gamma_{c1} - \gamma_1|$), which is fulfilled when $\gamma_1 = 9\text{ Hz}$, two zeros condense at a real frequency. Based on the above analyses, it is easy to give a feasible configuration of system parameters in the practical design satisfying the critical conditions of CPA EP. Under this configuration, we showcase the simulated reflection spectra $R_{1,2}$ and transmission spectra $T_{1,2}$ in Figure 2(c) and Figure 2(d) when the incident wave is emitted only from port 1 and port 3, respectively. Simulated and theoretical results agree well with each other, with both showing that $R_{1,2}$ and $T_{1,2}$ are almost equal at the resonant frequency $\omega_0$, except for the slight discrepancy between $R_2$ and $T_2$ caused by the backscattering effects in the cavity modes [36].

***Experimental observation of acoustic CPA EP.*** - Based on the theoretical analysis above, achieving CPA EP necessitates imbalanced internal losses between the two resonator cavities. However, since both resonant cavities have identical structures, their inherent losses are approximately equal. To realize the imbalanced losses which are crucial for the experimental observation of CPA EP, we introduce an active acoustic unit to flexibly modulate the non-Hermiticity while keeping the physical dimension of the cavity, as shown in Figure 3(a). The active non-Hermitian modulation unit comprises a transmitter, a receiver, and a feedback circuit installed at the top of each cavity. The amplitude and phase of the emission are precisely controlled according to the signal measured by the receiver. Gain/loss is introduced when the emission of the transmitter is in-phase/anti-phase with respect to the sound at the top of the cavity, respectively. Due to visco-thermal wall loss, there is an inherent background loss $\gamma_0$ in the acoustic resonant cavity. Note that $\gamma_0 = 15$Hz exceeds one of the intrinsic losses $\gamma_{1,2}$ ($\gamma_1 = 9$Hz and $\gamma_2 = 31$Hz) predicted under CPA EP, i.e., $\gamma_1 < \gamma_0 < \gamma_2$. A small amount of gain is added to the cavity A by an in-phase emission of the top transmitter, and the loss is injected in the cavity B using an anti-phase emission. By judiciously adjusting the active acoustic components, we can obtain the desired non-Hermiticity in the cavities to meet the predicted conditions for CPA EP. When the plane wave is emitted at port 1 or 3 only and the porous materials are added at ports 2 and 4 for avoiding the undesirable reflections. The reflection and transmission spectra $R_{1,2}$ and $T_{1,2}$ are measured, which are illustrated in Figure 3(b) and Figure 3(c). The measured spectra show good consistency with the simulated ones, where only a slight deviation occurs owing to unavoidable fabrication errors. This indicates that the current configuration satisfies the critical conditions of the CPA EP.

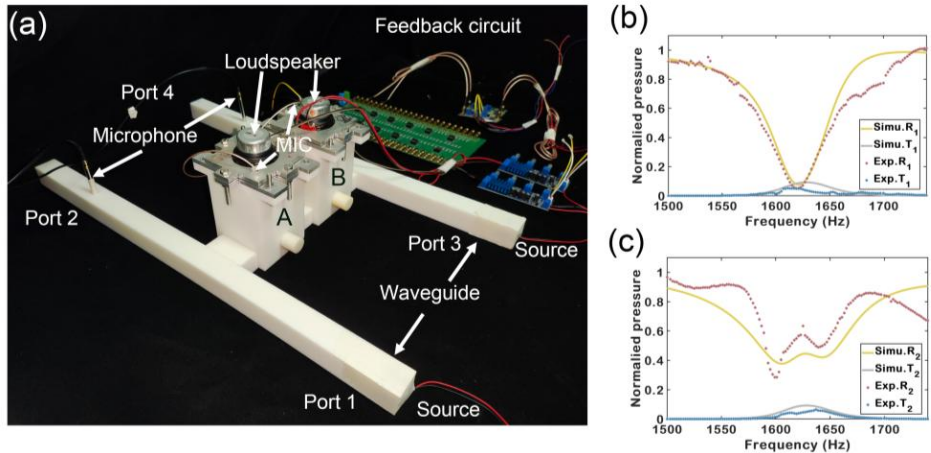

FIG.3. (a) Photograph of the experimental setup. (b) Experimentally measured reflection and transmission spectra by exciting port 1 only. For accurate estimation of the reflection and transmission coefficients, we normalize the acoustic pressure measured in the system with respect to that in an empty waveguide. (c) Experimentally measured spectra of reflection and transmission by exciting port 3 only.

Considering that the coherent absorption behavior is sensitive to the relative phase $\Delta\varphi$ and amplitude ratio $p_0$ of incoming acoustic wave $p_{i2}$ with respect to the other one $p_{i1}$, we inspect the total output power versus the input relative phase and amplitude ratio in Figure 4(a). Perfect absorption can be achieved when incoming acoustic waves match the eigenvector $v = [1, -i]^T$, which is expected to be observed at the minimum point on the output surface in Figure 4(a). In Figure 4(b), we further depict the relationship between the output spectrum and relative phase under equal incident amplitude. The output exhibits a sinusoidal pattern over the phase range and the expected perfect absorption is obtained with relative phase of -0.5π, consistent with theoretical predictions. This highlights the strong dependence of absorption on the input conditions due to the extreme sensitivity of CPA EP.

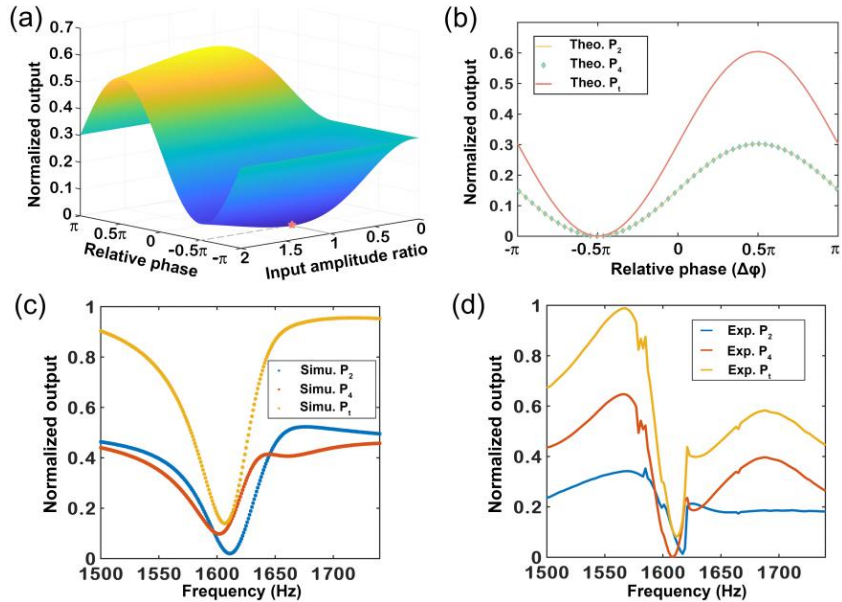

FIG.4. (a) Theoretical total output power normalized to the total input power at the zero detuning ($\delta = \omega - \omega_0 = 0$) as a function of the input amplitude ratio and relative phase. The total output power $P_t$ is the sum of the output $P_2$ from port 2 and output $P_4$ from port 4. (b) Theoretical output spectra as a function of the relative phase under equal incident amplitude. (c) and (d) Simulated and experimental spectra of the output power at the CPA EP.

In view of the dependence analysis on the incident wavefront, we properly tune the incoming waves generated by the transmitters with correct input power and relative phase, and perfect absorption is expected to be achieved due to a combination effect of interference and dissipation. The reflected part of the incident wave from port 1 interferes destructively with the transmitted part of the incident one from port 3, and vice versa, and therefore the radiation is trapped in an interference pattern within the lossy system and lost entirely to dissipation. Experimental results displayed in Figure 4(c) and Figure 4(d), in good agreement with the theoretical and simulated ones, show nearly perfect absorption at the resonant frequency, with negligible errors attributed to inevitable fluctuations of the incident wave phases and inaccuracies in tuning system non-Hermiticity via the electric circuit.

*Conclusions.* – In summary, we theoretically present an acoustic non-Hermitian scattering system composed of two-channel waveguides coupled to lossy resonant cavities to observe CPA EP. As a practical implementation of this system, a compact metamaterial-based model is proposed which allows independent modulation of the system parameters, where the precise adjustment of non-Hermiticity is enabled by the introduction of active acoustic components. We experimentally validate the occurrence of CPA EP, with measured results closely aligning with theoretical predictions and simulations, demonstrating strong absorption at the expected real frequency and extreme sensitivity to the phase variations of incoming acoustic waves. Our work enriches the non-Hermitian physics in classical wave systems and provides an important platform for investigation of intriguing phenomena occurring at EPs, which opens avenue for the design and application of singularity-based devices.

**Acknowledgments:**
This work was supported by the National Key R&D Program of China (Grant Nos. 2022YFA1404402), the National Natural Science Foundation of China (Grant Nos. 12174190), High-Performance Computing Center of Collaborative Innovation Center of Advanced Microstructures and A Project Funded by the Priority Academic Program Development of Jiangsu Higher Education Institutions. J. Christensen acknowledges support from the Spanish Ministry of Science and Innovation through a Consolidación Investigadora grant (CNS2022-135706).